\documentclass[aps,pre,showpacs,preprint,amsmath,amsfonts]{revtex4}

\input psfig.sty

\begin{document}

\title{Spectral method for the time-dependent Gross-Pitaevskii
  equation with a harmonic trap}

 \author{Claude M. Dion}
 \affiliation{CERMICS, \'{E}cole Nationale des Ponts et Chauss\'{e}es, 6 \& 8,
 avenue Blaise Pascal, Cit\'{e} Descartes, Champs-sur-Marne, 77455
 Marne-la-Vall\'{e}e, France}
 \author{Eric Canc\`{e}s}
 \affiliation{CERMICS, \'{E}cole Nationale des Ponts et Chauss\'{e}es, 6 \& 8,
 avenue Blaise Pascal, Cit\'{e} Descartes, Champs-sur-Marne, 77455
 Marne-la-Vall\'{e}e, France}

\date{\today} 

\begin{abstract}
We study the numerical resolution of the time-dependent Gross-Pitaevskii 
equation, a non-linear Schr\"{o}dinger equation used to simulate the dynamics of 
Bose-Einstein condensates.  Considering condensates trapped in harmonic 
potentials, we present an efficient algorithm by making use of a spectral 
Galerkin method, using a basis set of harmonic oscillator functions, and the 
Gauss-Hermite quadrature.  We apply this algorithm to the simulation of 
condensate breathing and scissors modes.
\end{abstract}


\maketitle

\section{Introduction}

The experimental realization of Bose-Einstein 
condensation~\cite{bec:anderson95,bec:bradley95,bec:davis95} has prompted 
much work on the study of the dynamics of these condensates.  From the 
theoretical side, many interesting results have been obtained using the 
Gross-Pitaevskii equation 
(GPE)~\cite{gp:gross61,gp:pitaevskii61,bec:dalfovo99},
\begin{equation}
i \hbar \frac{\partial \Psi}{\partial t} = \left[ -
    \frac{\hbar^{2}}{2m} \nabla^{2} + V_{\mathrm{ext}} + \frac{4 \pi
    \hbar^{2} a N}{m} \left| \Psi \right|^{2} \right] \Psi,
\label{eq:gp}
\end{equation}
with the normalization condition $\left\| \Psi(t) \right\|_{L^{2}} = 1\
\forall t$, to describe the order parameter $\Psi$ (also called the
\emph{condensate wave function}) of $N$ condensed bosons of mass $m$,
interacting via a contact potential described by the scattering length
$a$, and eventually confined by an external potential
$V_{\mathrm{ext}}$.  Even though the Gross-Pitaevskii equation is based
on the approximation that all bosons are in the condensed phase ($T =
0$~K), direct comparison between theoretical and experimental results
have shown that, in many cases, solutions of the GPE contain the
essential physics of the underlying
phenomena~\cite{bec:roehrl97,bec:bongs99,bec:matthews99,bec:modugno00}.
This non-linear Schr\"{o}dinger equation (NLSE) has been used, in its
time-dependent form, to investigate many aspects of the dynamics of
Bose-condensed gas, such as the formation of
vortices~\cite{bec:martikainen01}, the interference between
condensates~\cite{bec:hoston96}, of the possibility of creating atom
lasers~\cite{bec:ballagh97,bec:drummond00}, to mention only a few.

Most of theses and other numerical studies of the time-dependent GPE
are based on grid methods, i.e., discretize the spatial coordinates on
a grid of points, the resulting differential equation being usually
solved by Crank-Nicholson or split-operator Fourier methods (see,
e.g.,
Refs.~\cite{gp:ruprecht95,gp:adhikari00a,bec:jackson98,gp:baer00}).
We must point out that, while much care must be taken in solving
Eq.~(\ref{eq:gp}) because of the non-linearity, we find, to our
dismay, that many authors give results calculated with the
time-dependent GPE without even specifying what method they have used
for their numerical simulation.

In this article, we wish to focus our attention on the case where the 
Bose-Einstein condensate is in a (possibly anisotropic) harmonic trap, 
i.e.,
\begin{equation}
V_{\mathrm{ext}}(X,Y,Z) = \frac{1}{2} m \left( \omega_x^2 X^2 +
 \omega_y^2 Y^2 +  \omega_z^2 Z^2 \right)
\end{equation}
which is the case for most experimental
set-ups~\cite{af:tollett95,af:petrich95}.  The method we propose is
based on the spectral decomposition of $\Psi$ on a basis of
harmonic-oscillator wave functions.  In such a representation, the
kinetic + trapping potential part of the Hamiltonian is diagonal.  The
non-linear part is computed by forward and backward transformations
from the spectral to a grid representation.  By judicious use of the
Gauss-Hermite quadrature, this can lead to an algorithm that is more
efficient than those based on grid methods.  Although this is akin to
DVR methods based on Hermite polynomials, which have been successfully
used for the time-independent and time-dependent
GPE~\cite{bec:schneider99a,bec:feder00}, our method is distinct, since
our Hamiltonian is expressed in the spectral representation for both
the kinetic and potential operators.

We expose in Sec.~\ref{sec:spectr} our spectral method and the
resulting algorithm.  We then present different time evolution schemes
that can be used in combination with the spectral method.  We finally
give in Sec.~\ref{sec:results} some results that can be obtained from
the numerical simulation of the time-dependent GPE, namely the study
of condensate breathing and scissors modes.

\section{Space discretization} \label{sec:spectr}

To simplify the calculation, we will first rescale Eq.~(\ref{eq:gp}) in the 
three spatial dimensions $(X,Y,Z)$ and in time,
\begin{subequations}
\begin{eqnarray}
X & = & \left( \frac{\hbar}{m \omega_{x}} \right)^{1/2} x \\
Y & = & \left( \frac{\hbar}{m \omega_{y}} \right)^{1/2} y \\
Z & = & \left( \frac{\hbar}{m \omega_{z}} \right)^{1/2} z \\
t & = & \frac{1}{\omega_{z}} \tau. \label{eq:rescalet}
\end{eqnarray}
\end{subequations}
We also introduce a new wave function $\psi$ defined as
$$
\Psi(t,X,Y,Z) = A \psi(\tau,x,y,z)
$$
and, considering the normalization condition
$$
\int_{\mathbb{R}^{3}} \left| \Psi(t,X,Y,Z) \right|^{2} dX dY dZ = 1\
\forall t,
$$
we choose
$$
A  =  \left( \frac{m}{\hbar} \right)^{3/4} \left( \omega_{x}
    \omega_{y} \omega_{z} \right)^{1/4}
$$
such that
$$
\int_{\mathbb{R}^{3}} \left| \psi(\tau,x,y,z) \right|^{2} dx dy dz = 1.
$$
The Gross-Pitaevskii equation therefore becomes
\begin{equation}
i \frac{\partial \psi}{\partial \tau} = \left[
  \frac{\omega_{x}}{\omega_{z}} \left( -\frac{1}{2} 
  \nabla_{x}^{2} + \frac{x^{2}}{2} \right) +
  \frac{\omega_{y}}{\omega_{z}} \left( -\frac{1}{2} \nabla_{y}^{2} +
  \frac{y^{2}}{2} \right) + \left(
  -\frac{1}{2} \nabla_{z}^{2} + \frac{z^{2}}{2} \right) +
\lambda \left| \psi \right|^{2} \right] \psi
\label{eq:gp3D}
\end{equation}
with
\begin{equation}
\lambda = 4 \pi a N \left(\frac{m}{\hbar} \frac{\omega_{x}
    \omega_{y}}{\omega_{z}} \right)^{1/2}.
\label{eq:lambda}
\end{equation}
Coordinate $z$ should be chosen such that $\omega_{z}$ is the greatest
of the three frequencies [this is related to the arbitrary choice of
the scaling factor in Eq.~(\ref{eq:rescalet})].

As all the physical parameters have been absorbed in the non-linear 
parameter $\lambda$, calculations with the same $\lambda$ can correspond to 
results for different species, but in diverse experimental conditions.  We 
can define acceptable lower and upper bounds for $\lambda$ by considering 
the effective range of the different physical parameters.  Considering only 
cases where the interparticle interaction is repulsive, i.e., $a > 0$ and 
therefore $\lambda > 0$, at the lower end we can consider a small 
$^{4}$He$^{*}$ condensate ($m=4.0 \mbox{ a.m.u.}$, $a=302 \mbox{ 
a.u.}$~\cite{bec:pereira01}) of $N = 10^{3}$ atoms in a highly anisotropic 
$\omega_{x}\omega_{y}/\omega_{z} = 2\pi \times 10^{-1} \mbox{ Hz}$ trap, 
giving $\lambda \approx 1.3$, while for a bigger $N \sim 10^{6}$ condensate of 
heavy atoms such as $^{87}$Rb ($m=86.9 \mbox{ a.m.u.}$, $a = 106 
\mbox{ a.u.}$~\cite{Rb2:roberts98}), $\lambda$ can reach $10^{5}$ for 
isotropic traps.  In the following, we will restrict our study to $\lambda$ 
in the range 1--$10^{3}$, considering that the Thomas-Fermi approximation 
can be used for greater values of $\lambda$~\cite{bec:schneider99a}.

\subsection{The spectral Galerkin method in 1D} \label{sec:Galerkin1D}

For pedagogical purposes, we first explain our numerical method on
the simple case of the one-dimensional NLSE 
\begin{equation} \label{eq:GP1D}
i \frac{\partial \psi}{\partial t}(t,x) = H_0 \psi(t,x) + \lambda
|\psi(t,x)|^2 \psi(t,x) 
\end{equation}
with
$$
H_0 = - \frac{1}{2} \frac{\partial^2}{\partial x^2} +
  \frac 1 2 x^2.
$$
Extensions to the three-dimensional case will be detailed in the next
section.  

Denoting by $\psi(t)$ the function $x \mapsto \psi(t,x)$,
it can be proven \cite{nls:cazenave96} that if 
\[
\psi_0 \in \mathcal{X}:= \left\{ \chi \in L^2(\mathbb{R}), \quad \int_\mathbb{R}
      \left|\frac{\partial 
      \chi}{\partial x}\right|^2 < + \infty, \quad \int_\mathbb{R} x^2
  |\chi(x)|^2 dx  < + \infty \right\}, 
\]
Eq.~(\ref{eq:GP1D})
with initial condition $\psi_0$ has a unique solution in 
$C^0([0,+\infty[,\mathcal{X}) \cap C^1([0,+\infty[,L^2(\mathbb{R}))$ and that both the
$L^2$ norm $$
\| \psi(t) \|_{L^2} =  \left[ \int_{\mathbb{R}} |\psi(t,x)|^2 dx
\right]^{1/2} 
$$
and the energy
$$
E = (H_0 \psi(t),\psi(t)) + \frac \lambda 2 \int_{\mathbb{R}} |\psi(t,x)|^4  dx 
$$
are conserved by the dynamics. A variational formulation of
Eq.~(\ref{eq:GP1D}), supplemented by the initial condition
$\psi(t=0)=\psi_0$ where $\psi_0 \in \mathcal{X}$, reads 
\begin{equation} \label{eq:GP1DVar}
\left\{ \begin{array}{l} \mbox{Search $\psi \in C^0([0,T],\mathcal{X}) \cap
      C^1([0,T],L^2(\mathbb{R}))$ such that} \\
\displaystyle{\forall \chi \in \mathcal{X}, \quad i \frac{d}{dt}
  (\psi(t),\chi)  = (H_0 \psi(t),\chi) + \lambda
  (|\psi(t)|^2 \psi(t),\chi) } \\
\psi(0) = \psi_0. \end{array} \right. 
\end{equation}
Numerical solutions can then be obtained by approximating 
problem~(\ref{eq:GP1DVar}) with a Galerkin method: a \emph{finite} 
dimensional subspace $\mathcal{X}_N$ of the \emph{infinite} dimensional 
vector space $\mathcal{X}$ being given, we consider
\begin{equation} \label{eq:GP1DGalerkin}
\left\{ \begin{array}{l} \mbox{Search $\psi_N \in C^1([0,T],\mathcal{X}_N)$
      such that} \\ 
\displaystyle{\forall \chi_N \in \mathcal{X}_N, \quad i \frac{d}{dt}
  (\psi_N(t),\chi_N)  = (H_0 \psi_N(t),\chi_N) + \lambda
  (|\psi_N(t)|^2 \psi_N(t),\chi_N) } \\
\psi_N(0) = \psi_0. \end{array} \right. 
\end{equation}
Denoting by $(\phi_0,\ldots,\phi_N)$ an orthonormal basis of $X_N$ for
the $L^2$ scalar product and by $C(t)=(c_n(t))_{0 \le n \le N}$ the
vector of $\mathbb{C}^{N+1}$ collecting the coefficients of
$\psi_N(t)$ in the basis $(\phi_0,\ldots,\phi_N)$, i.e.,
$$
\psi_N(t,x) = \sum_{n=0}^N c_n(t) \phi_n(x), 
$$
problem~(\ref{eq:GP1DGalerkin}) can be reformulated as 
\begin{equation}
\left\{ \begin{array}{l} \mbox{Search $C \in C^1([0,T],\mathbb{C}^{N+1})$
      such that} \\ 
\displaystyle{ i \frac{dC}{dt}(t) = h C(t) + \lambda \, F(C(t)) } \\
C(0) = C_0 \end{array} \right. 
\end{equation}
where $C_0$ are the coefficients of $\psi_0$ and $h$ the matrix of $H_0$
in the basis $(\phi_0,\ldots,\phi_N)$
$$
[C_0]_n = (\psi_0,\phi_n)_{L^2}, \qquad h_{nm} = (H_0 \phi_m, \phi_n),
$$
and where the function $F$ is defined by 
\begin{equation}
F(C)_n = \sum_{k,l,m=0}^N I_{klmn} c_{k}^{*} c_l c_m, 
\label{eq:direct}
\end{equation}
with
$$
\quad I_{klmn} = \int_{\mathbb{R}} \phi_{k}^{*} \phi_l \phi_m \phi_{n}^{*}.
$$
The efficiency of a direct
implementation~\cite{bec:edwards96a,bec:edwards96b} of the Galerkin
method described above is very poor: the calculation of the integrals
$I_{klmn}$ (which can be precomputed if the basis is small
enough that the integrals can be stored in memory) scales as
$O(N^4N_p)$ where $N_p$ is the number of grid points of the quadrature
method, and the computation cost for one evaluation of the function
$F$ scales as $N^4$ [for each of the $N$ coefficients, $O(N^3)$
operations are needed].

Our aim is to show that the Galerkin method becomes very efficient if
$(\phi_0, \ldots, \phi_N)$ are the $N+1$ lowest eigenmodes of the
harmonic oscillator $H_0$. In this case, indeed, the vector $F(C)$ can
be computed \emph{exactly} (up to round-off errors) in $O(N^2)$
operations. Let us recall that the eigenmodes $(\phi_n)_{n \in
  \mathbb{N}}$ of $H_0$ read
$$
\phi_n(x) = \mathcal{H}_n(x) e^{-x^2/2},
$$
where $\mathcal{H}_n(x)$ is the $n$-th Hermite
polynomial~\cite{cohen-tannoudji92}, and that they satisfy
$$
H_0 \phi_n = E_n \phi_n, \quad \mbox{with} \quad E_n=n+\frac 1 2.
$$
In such a basis, the matrix $h$ is therefore diagonal:
$h=\mbox{Diag}(E_0,\ldots,E_n)$. In addition, for any $C \in \mathbb{C}^{N+1}$,
one has
\begin{equation} \label{eq:defF}
F(C)_n = \int_{\mathbb{R}} |\psi(x)|^2 \psi(x)  \phi_n(x) dx,
\end{equation}
where $\psi(x) = \sum_{n=0}^N c_n \phi_n(x)$.  The key point is now that for 
any $n \le N$ the integrand in~(\ref{eq:defF}) is of the form $Q(x) \, 
e^{-2x^2}$, where $Q(x)$ is a polynomial of degree lower or equal to $4N$; 
each of the $N+1$ integrals can therefore be computed \emph{exactly} with a 
Gauss-Hermite quadrature formula involving $2N$ Gauss 
points~\cite{math:stroud66}.  More precisely, we have, for any polynomial $Q$ 
of degree lower or equal to $4N$,
$$
\int_{-\infty}^{+\infty} Q(x) e^{-x^2} dx = \sum_{k=1}^{2N+1} w_k Q(x_k)
$$
where $\left\{ x_k \right\}$ are the roots of the Hermite
polynomial $\mathcal{H}_{2N+1}$ and where $\left\{ w_k \right\}$ are
convenient weights~\cite{hmf:davis2}.  By a change of variable in
integral (\ref{eq:defF}), it follows that
$$
F(C)_n =  \sum_{k=1}^{2N+1} 
\left( \frac{w_k e^{x_k^2}}{\sqrt{2}} \right) \left| 
\psi \left(x_k/\sqrt{2}\right) \right|^2 \psi \left(x_k/\sqrt{2}\right) 
\phi_n  \left(x_k/\sqrt{2}\right) .
$$
Spectral Galerkin methods are usually not very
efficient~\cite{math:maday00}; but they can be in the specific case of
the NLSE we are interested in because of the special form of the
nonlinearity.

Let us now denote by $P \in \mathcal{M}(N+1,2N+1)$ the matrix collecting
the values of the basis functions $(\phi_n)_{0 \le n \le N}$ at the
Gauss points $(x_k)_{1 \le k \le 2N+1}$:
$$
P_{nk} = \phi_n \left( x_k / \sqrt{2} \right),
$$
and by $\tilde{w}_k = w_k e^{x_k^2}/\sqrt{2}$. 
An efficient algorithm for the computation of $F(C)$ for a given $C \in
\mathbb{C}^{N+1}$  reads:
\begin{enumerate}
\item Compute the vector $\Psi \in \mathbb{C}^{2N+1}$ defined by 
$$
\Psi = P^T \cdot C.  
$$
\item Compute the vector $\Xi \in \mathbb{C}^{2N+1}$ coefficient by coefficient
  along formula
$$
\Xi_k = \tilde w_k |\Psi_k|^2 \Psi_k.
$$
\item Compute
$$
F(C) = P \cdot \Xi. 
$$
\end{enumerate} 
The vectors $C$ and $\Psi$ are the representation of the wave function  
$\psi$ in the spectral basis $\left\{ \phi_n \right\}_{0 \le n \le
  N}$ and in real space (at the $2N+1$ Gauss  
points $\left\{ x_k/\sqrt{2} \right\}$), respectively. Steps~1 and~3 of
the above  
algorithm scale quadratically in $N$ (these are matrix-vector products),
and step~2 scales linearly in $N$.
We therefore end up with an algorithmic complexity in $O(N^2)$.

In practice, the function $C \mapsto F(C)$ is called one or several times at 
each time step; of course, the matrix $P$ as well as the weights $\tilde 
w_k$ can be precomputed once and for all and stored in memory.

\subsection{The spectral-Galerkin method in 3D} \label{sec:Galerkin3D}

Let us now turn to the 3D setting and consider the rescaled equation
\begin{equation} \label{eq:GP3D}
i \frac{\partial \psi}{\partial t}(t,x,y,z) = \left[
  \frac{\omega_x}{\omega_z} H_0(x)  +
\frac{\omega_y}{\omega_z} H_0(y)  +
 H_0(z) \right] \psi(t,x,y,z) 
+ \lambda |\psi(t,x,y,z)|^2 \psi(t,x,y,z) 
\end{equation}
with
$$
H_0(x) = - \frac 1 2 \frac{\partial^2}{\partial x^2} +
  \frac 1 2 x^2, \quad H_0(y) = - \frac 1 2
  \frac{\partial^2}{\partial y^2} + \frac 1 2 y^2,  \quad H_0(z) =
- \frac 1 2 \frac{\partial^2}{\partial z^2} + \frac 1 2
  z^2.
$$
For $\lambda \ge 0$, a global-in-time existence and uniqueness result is
available for Eq.~(\ref{eq:GP3D}) with initial condition 
$\psi(t=0) = \psi_0$ and
\[
\psi_0 \in \mathcal{X} = \left\{ \chi \in L^2(\mathbb{R}^3), \quad \nabla \chi \in
  \left( L^2(\mathbb{R}^3) \right)^3, \quad \left(x^2+y^2+z^2\right)^{1/2} \chi
  \in L^2(\mathbb{R}^3) \right\}.
\]
On the other hand, it is well known that finite-time blow-up may be observed 
for $\lambda < 0$ and for some initial conditions~\cite{nls:cazenave96}.  As 
stated above, we focus here on the case where $\lambda \ge 0$.

Following the same lines as in the Sec.~\ref{sec:Galerkin1D}, the 
approximated wave function $\psi_N(t)$ is expended on the spectral tensor 
basis set
$$
(\phi_{n_x}(x) \;  \phi_{n_y}(y) \; \phi_{n_z}(z))_
{0 \le n_x \le N_x, \; 0 \le n_y \le N_y, \; 0 \le n_z \le N_z}.
$$
One therefore has
\begin{equation}
\psi_N(t,x,y,z) = \sum_{n_x=0}^{N_x} \sum_{n_y=0}^{N_y} \sum_{n_z=0}^{N_z}
c_{n_x n_y n_z}(t)  \phi_{n_x}(x)   \phi_{n_y}(y)  \phi_{n_z}(z).
\label{eq:basis}
\end{equation}
The equation satisfied by the three index tensor $C = [c_{n_x n_y n_z}]$
in the Galerkin approximation formally has the same expression as in 1D, 
$$
i \frac{dC}{dt}(t) = h C(t) + \lambda F(C(t)),
$$
the linear operator $h$ now being defined by
$$
[hC]_{n_x n_y n_z} = E_{n_x n_y n_z} c_{n_x n_y n_z}
$$
with
$$
E_{n_xn_yn_z} = \frac{\omega_{x}}{\omega_{z}} \left( n_x + \frac{1}{2} 
\right) + \frac{\omega_{y}}{\omega_{z}} \left( n_y + \frac{1}{2} \right) + 
\left( n_z + \frac{1}{2} \right),
$$
and the non-linear function $F(C)$ by
$$
[F(C)]_{n_xn_yn_z} = \int_{\mathbb{R}^3} |\psi(x,y,z)|^2 \psi(x,y,z) 
\phi_{n_x}(x) \phi_{n_y}(y) \phi_{n_z}(z) dx dy dz,
$$
where $\psi(x,y,z)$ is given by Eq.~(\ref{eq:basis}).

Let us denote by $\left\{ x_k \right\}_{1 \le k \le 2N_x+1}$, $\left\{ y_k 
\right\}_{1 \le k \le 2N_y+1}$, $\left\{ z_k \right\}_{1 \le k \le 2N_z+1}$ 
the roots of the Hermite polynomials $\mathcal{H}_{2N_x+1}$, 
$\mathcal{H}_{2N_y+1}$, $\mathcal{H}_{2N_z+1}$ and $\left\{ w^x_k 
\right\}_{1 \le k \le 2N_x+1}$, $\left\{ w^y_k \right\}_{1 \le k \le 
2N_y+1}$, $\left\{ w^z_k \right\}_{1 \le k \le 2N_z+1}$ the associated 
summation weights.  Let us also introduce the matrices $P_x \in 
\mathcal{M}(N_x+1,2N_x+1)$, $P_y \in \mathcal{M}(N_y+1,2N_y+1)$, $P_z \in 
\mathcal{M}(N_z+1,2N_z+1)$ defined by
$$
[P_x]_{n_xk_x} = \phi_{n_x}(x_{k_x}/\sqrt{2}), \qquad 
[P_y]_{n_yk_y} = \phi_{n_y}(y_{k_y}/\sqrt{2}), \qquad 
[P_z]_{n_zk_z} = \phi_{n_z}(z_{k_z}/\sqrt{2}),
$$
and the weights 
$$
\tilde w^x_{k_x} = \frac{w^x_{k_x} e^{x_{k_x}^2}}{\sqrt 2}, \qquad
\tilde w^y_{k_z} = \frac{w^y_{k_y} e^{y_{k_y}^2}}{\sqrt 2}, \qquad
\tilde w^z_{k_z} = \frac{w^z_{k_z} e^{z_{k_z}^2}}{\sqrt 2}.
$$
The following algorithm for the computation of $F(C)$ scales in
$O(N N_x N_y N_z)$ where $N=\max(N_x,N_y,N_z)$: \\
\begin{tabular}{rll}
1. & Set $\Psi^{SSS} = C$ \\
2. & Compute $\displaystyle{\Psi^{SSR}_{n_xn_yk_z} =
    \sum_{n_z=0}^{N_z} [P_z]_{n_zk_z} \Psi^{SSS}_{n_xn_yn_z}}$
& $O(N_xN_yN_z^2)$ operations \\
3. & Compute $\displaystyle{\Psi^{SRR}_{n_xk_yk_z} =
    \sum_{n_y=0}^{N_y} [P_y]_{n_yk_y} \Psi^{SSR}_{n_xn_yk_z}}$ &
$O(N_xN_y^2N_z)$ operations \\
4. & Compute $\displaystyle{\Psi^{RRR}_{k_xk_yk_z} =
    \sum_{n_x=0}^{N_x} [P_x]_{n_xk_x} \Psi^{SRR}_{n_xk_yk_z}}$ &
  $O(N_x^2N_yN_z)$ operations \\
5. & Compute $\displaystyle{\Xi^{RRR}_{k_xk_yk_z} =
\tilde w^x_{k_x} \tilde w^y_{k_y}  \tilde w^z_{k_z}
|\Psi^{RRR}_{k_xk_yk_z}|^2 \Psi^{RRR}_{k_xk_yk_z}}$ \quad \quad &
$O(N_xN_yN_z)$ operations \\ 
6. & Compute $\displaystyle{\Xi^{RRS}_{k_xk_yn_z} =
    \sum_{k_z=1}^{2N_z+1} [P_z]_{n_zk_z} \Xi^{RRR}_{k_xk_yk_z}}$ &
  $O(N_xN_yN_z^2)$ operations \\
7. & Compute $\displaystyle{\Xi^{RSS}_{k_xn_yn_z} =
    \sum_{k_y=1}^{2N_y+1} [P_y]_{n_yk_y} \Xi^{RRS}_{k_xk_yn_z}}$ &
  $O(N_xN_y^2N_z)$ operations \\
8. & Compute $\displaystyle{\Xi^{SSS}_{n_xn_yn_z} =
    \sum_{k_x=1}^{2N_x+1} [P_x]_{n_xk_x} \Xi^{RSS}_{k_xn_yn_z}}$ &
  $O(N_x^2N_yN_z)$ operations \\
9. & Set $F(C) = \Xi^{SSS}$.
\end{tabular} \\
In the above formulation, the superscripts $S$ and $R$ stand for
\emph{spectral} and \emph{real space} representations respectively.  In 
other words, steps 2--4 constitute the successive transform of the 
wave function from the spectral basis to a spatial representation on the 
series of points of the Gauss-Hermite quadrature.  The non-linear term of 
the Hamiltonian is then calculated in this spatial representation (step 5), 
while steps 6--8 correspond to the inverse transform back to the spectral 
basis.  It is this procedure of forward and backward transformation that 
allows us to obtain a much better scaling than the implementation of 
Eq.~(\ref{eq:direct}).

The scaling of the above algorithm [$O(N^4)$ if $N_x=N_y=N_z$] has to be 
compared with the scaling of FFT based algorithms which scale in $O(N_p^3 
\log(N_p))$ where $N_p$ is the number of grid points per direction.  The 
main interest of the spectral method is that for a similar accuracy, the 
number of spectral basis functions per direction (here denoted by $N$) can 
usually be chosen much smaller than the number $N_p$ of grid points per 
direction.  This is especially true when the problem considered displays a 
symmetry in one or more of the directions, in which case the basis set used 
in the Galerkin approximation Eq.~(\ref{eq:basis}) can be restricted to even 
harmonic oscillator functions (in the corresponding direction).  We will 
come back on this important feature of the spectral method in 
Section~\ref{sec:results}.

\subsection{Exploiting spherical or cylindrical symmetry}

When $\omega_x = \omega_y = \omega_z$ the one-particle Hamiltonian
possesses spherical symmetry. If the initial condition $\psi_0 = \psi(t=0)$
has the same symmetry, then the wave function $\psi(t)$ is spherical
symmetric for any $t > 0$: $\psi(t,x,y,z) = \psi(t,r)$ where
$r=(x^2+y^2+z^2)^{1/2}$, is the radial coordinate. Eq.~(\ref{eq:gp3D})
leads to the effective 1D dynamics
\begin{equation}
i \frac{\partial \psi}{\partial t} = \left[ -\frac{1}{2r^{2}} 
  \frac{\partial}{\partial r} \left( r^{2} \frac{\partial}{\partial 
  r} \right) + \frac{r^{2}}{2} + \lambda \left| \psi \right|^{2} 
  \right] \psi.
\label{eq:gp1D}
\end{equation}
Let us now define the function 
$$
\chi(t,r) =  \left| \begin{array}{lll} 
\displaystyle{ \sqrt{2\pi} \; r \psi(t,r) } & \quad & \mbox{ if } r > 0 \\
\displaystyle{ - \sqrt{2\pi} \; r \psi(t,-r) } & \quad & \mbox{ if } r < 0 .
\end{array} \right.
$$
It is easy to check that $\chi$ actually satisfies the 1D
NLSE
$$
i  \frac{\partial \chi}{\partial t} = H_0 \chi  + \lambda
\frac{|\chi|^2}{2 \pi r^2} \chi.
$$
Besides, for any $t > 0$ the function $\chi(t):\ r \mapsto
\chi(t,r)$ is odd and belongs to $L^2(\mathbb{R})$ since 
$$
\int_{-\infty}^{+\infty} |\chi(t,r)|^2 \, dr =  \int_0^{+\infty}
4 \pi \, r^2 |\psi(t,r)|^2 \, dr = 1. 
$$
It can thus be expanded on the \emph{odd} modes of the harmonic
oscillator:
$$
\chi(t,r) = \sum_{n=0}^{+\infty} c_n(t) \phi_{2n+1}(r).
$$
A spectral Galerkin approximation can now be used. The vector $C(t) \in
\mathbb{C}^{N+1}$ collecting the coefficients $(c_k(t))_{0 \le k \le N}$ of the
approximated wave function  
$$
\chi_N(t,r) = \sum_{n=0}^{N} c_n(t) \phi_{2n+1}(r)
$$
obeys once again a dynamics of the form
$$
i \frac{dC}{dt}(t) = hC(t) + \lambda F(C(t)).
$$
Here 
$$
h = \mbox{Diag}(E_{2n+1}), \qquad \mbox{with} \qquad 
E_{2n+1} = 2n + \frac 3 2,
$$
and
$$
[F(C)]_n = \int_{\mathbb{R}} \frac{|\chi(r)|^2}{2 \pi r^2} \chi(r) 
\phi_{2n+1}(r) dr,
$$
where $\chi(r) = \sum_{n=0}^{N} c_n \phi_{2n+1}(r)$. As
for any $0 \le n \le N$, $\phi_{2n+1}(r) = r P_{2n}(r) e^{-r^2/2}$
where $P_{2n}$ is a polynomial of degree equal to $2n$, it follows that
the above integrals can be computed exactly with $4N$ Gauss points.

Let us now turn to the cylindrical symmetry when (for instance) $\omega_x =
\omega_y$ and when the initial data reads $\psi_0(x,y,z)= \psi_0(r,z)$
with $r = (x^2+y^2)^{1/2}$. In this case, the cylindrical symmetry is
preserved by the dynamics so that for any $t > 0$, $\psi(t,x,y,z) =
\psi(t,r,z)$ and the time evolution of $\psi(t,r,z)$ is then governed by
the 2D equation 
\begin{equation}
i \frac{\partial \psi}{\partial t} = \left[ \frac{\omega_{x}}{\omega_{z}} 
\left( -\frac{1}{2r} \frac{\partial}{\partial r} \left( r 
\frac{\partial}{\partial r} \right) + \frac{r^{2}}{2} \right) + \left( 
-\frac{1}{2} \frac{\partial^{2}}{\partial z^{2}} + \frac{z^{2}}{2} \right) + 
\lambda \left| \psi \right|^{2} \right] \psi,
\label{eq:gp2D}
\end{equation}
set on the spatial domain $\mathbb{R}^+ \times \mathbb{R}$.
Defining a new function $\chi(t,r,z)$ on the space domain $\mathbb{R}^2$ by 
$$
\chi(t,r,z) = \left| \begin{array}{lll} 
\displaystyle{ \psi(t,r,z) } & \quad & \mbox{ if } r > 0 \\
\displaystyle{ \psi(t,-r,z) } & \quad & \mbox{ if } r < 0 
\end{array} \right.
$$
it occurs that $\chi$ satisfies 
\begin{equation}
i \frac{\partial \chi}{\partial t} = \left[ \frac{\omega_{x}}{\omega_{z}} 
\left( -\frac{1}{2} \frac{\partial^{2}}{\partial r^{2}} + \frac{r^{2}}{2} 
\right) + \left( -\frac{1}{2} \frac{\partial^{2}}{\partial z^{2}} + 
\frac{z^{2}}{2} \right) - \frac{\omega_{x}}{\omega_{z}} \frac 1 {2r} 
\frac{\partial}{\partial r}+ \lambda \left| \chi \right|^{2} \right] \chi
\label{eq:gp2Dho}
\end{equation}
on the space domain $\mathbb{R}^2$, and that, by construction, the function $r
\mapsto \chi(t,r,z)$ is even. A spectral Galerkin approximation is
obtained by expanding the wave function on the spectral tensor basis set
$$
(\phi_{2n_r}(r) \; \phi_{n_z}(z))_{0 \le n_r \le N_r, 0 \le n_z \le N_z}.
$$
The coefficients $(c_{n_rn_z})_{0 \le n_r \le N_r, 0 \le n_z \le N_z}$
of the expansion are solution of an equation of the same form as above,
$$
i \frac{dC}{dt}(t) = hC(t) + \lambda F(C(t)).
$$
The main difference is that in this case, the linear map $h$ takes
into account the operator $- \frac 1 {2r} \frac{\partial}{\partial
  r}$:
$$
[hC]_{n_rn_z} = \left[ \frac{\omega_{x}}{\omega_{z}} \left( n_r+ \frac 1 2 
\right) + \left( n_z+ \frac 1 2 \right) \right] C_{n_rn_z} - \frac{1}{2} 
\frac{\omega_{x}}{\omega_{z}} \sum_{m_r = 0}^{N_r} \left( \frac{1}{r} 
\frac{d \phi_{2m_r}}{d r} , \phi_{2n_r} \right)_{L^2} C_{m_rn_z}.
$$
Let us remark that the scalar product $\left( \frac 1 r \frac{d 
\phi_{2m_r}}{d r} , \phi_{n_r} \right)_{L^2}$ is well defined since the 
first derivative of $\phi_{2n_r}$ is of the form $r P_{2n_r}(r) e^{-r^2/2}$ 
where $P_{2n_r}$ is a polynomial of degree $2n_r$; in addition, it can be 
computed exactly by numerical integration with $2n_r$ Gauss points.  It is 
worth pointing out that the ``Hamiltonian'' in~(\ref{eq:gp2Dho}) is not 
self-adjoint because of the term $- \frac 1 {2r} \frac{\partial}{\partial 
r}$ and that the $L^2$ norm of $\chi(t)$ is not a conserved quantity; on the 
other hand, the $L^2$ norm of $\chi(t)$ {\em for the measure} $r\, dr \, dz$ 
is conserved.

\section{Time discretization} \label{sec:time}

When a spectral Galerkin method is used to discretize the space
variables, one ends up with a finite dimensional dynamical system of the
form
\begin{equation} \label{eq:EDO}
i \frac{dC}{dt}(t) = h C(t) + \lambda \, F(C(t)),
\end{equation}
with initial condition $C(t=0) = C_0$.  We then use a basic
fourth-order Runge-Kutta method~\cite{press92} to solve
Eqs.~(\ref{eq:EDO}).  Let us mention that, as the Hamiltonian
character of the NLSE is preserved by the spectral Galerkin
discretization, it would be possible to resort to symplectic
methods~\cite{gp:tang96}; such algorithms, which are particularly
advised for long time evolution, are however not tested in the present
work.

We will also use a grid method, based on the split-operator method, to serve 
as a benchmark for the spectral algorithm we have just detailed.  We recall 
below the main features of this approach.

The wave function at time $\tau + \Delta \tau$ can be obtained from
the wave function at $\tau$ according to
\begin{equation}
\psi(\tau + \Delta \tau) = \hat{U}(\tau, \tau + \Delta \tau) \psi(\tau), 
\end{equation}
with the propagator $\hat{U}(\tau, \tau + \Delta \tau)$ being expressed, for
sufficiently small intervals $\Delta \tau$, as
\begin{equation}
\hat{U}(\tau, \tau + \Delta \tau) = \exp \left[ -i H(\tau)
  \Delta \tau \right],
\end{equation}
where $H(\tau)$ is the Hamiltonian of Eq.~(\ref{eq:gp3D}). 
As the potential and non-linear components of the Gross-Pitaevskii
Hamiltonian do not commute with the kinetic operator, we apply the
split-operator method~\cite{fft:feit82} to obtain
\begin{equation}
\exp \left[ -i H(\tau) \Delta \tau \right] = \exp \left[ -i T \frac{\Delta 
\tau}{2} \right] \exp \left[ -i \left( V + \lambda \left| \psi \right|^{2} 
\right) \Delta \tau \right] \exp \left[ -i T \frac{\Delta \tau}{2} \right] + 
O(\Delta \tau^{3}),
\end{equation}
with $T$ the kinetic operator and $V$ the trapping potential.
The middle term is diagonal in position space, while the kinetic part
is diagonal in momentum space.  A Fast Fourier Transform is thus used
before application of the kinetic operator, followed by the inverse
transform.  Note that if the intermediate wave function at time $\tau
+ \Delta \tau$ is not needed, the two successive kinetic operators
half-steps can be combined.  From a previous study~\cite{gp:taha84},
it appears that the split-operator method is the fastest algorithm for
solving a NLSE on a grid.

\section{Results} \label{sec:results}

The first test we perform is the propagation of the ground stationary
state (obtained from the time-independent GPE solved by a method based
on the Optimal Damping
Algorithm~\cite{oda:cances00a,oda:cances00b,gp:cances02}), while
monitoring the value of the coefficients $c(\tau)$ of the
expansion~(\ref{eq:basis}).  For the spherically symmetric case, we
require that the relative error on the $c_0$ coefficient (which has
the largest absolute value) be inferior to $10^{-8}$, i.e., $\left|
  \left| c_{0}(\tau) \right|^{2} - \left| c_{0}(\tau=0) \right|^{2}
\right| /\left| c_{0}(\tau=0) \right|^{2} \leq 10^{-8}\ \forall \tau \in
[0,100]$.  This criterion also results in an absolute error of all
coefficients $\left| c_{n}(\tau) \right|^{2} - \left| c_{n}(\tau=0)
\right|^{2} \leq 10^{-8}$.  We have also checked that the phase of the
coefficients is correct, by calculating $\left| c_{n}(\tau) -
  c_{n}(\tau=0) e^{-i \mu \tau} \right|^{2}/\left|
  c_{n}(\tau)\right|^{2}$, where $\mu$ is the chemical potential of
the ground stationary state of the GPE~\cite{bec:dalfovo99}, and this
value indeed is less than $10^{-12}$.

In this 1D case, we need $N=20$ basis functions for $\lambda = 100$,
and the resulting time-step for the Runge-Kutta propagator is
$\Delta\tau = 0.005$.  If $\lambda = 1000$, the basis set used should
be slightly larger, $N=26$, with a smaller time step $\Delta\tau =
0.0025$ to insure that the above error criteria are met.  The
resulting propagation time up to $\tau=100$ is 8.9~s for $\lambda =
100$ (calculated on an Athlon 1.2~GHz processor running under Linux,
using the NAG Fortran 90 compiler at the \texttt{-O} level of
optimization) and 28.3~s for $\lambda = 1000$.  If we double the size
of the basis set, we get a CPU time of 32.9~s for $\lambda = 100$,
showing the expected $O(N^{2})$ scaling of the algorithm in 1D.

Comparing now with the grid method described in Sec.~\ref{sec:time},
we use $N_{p} = 64$ grid points in the range $-8 \leq r \leq 8$.  The
time step used is $\Delta\tau = 0.00025$, resulting in a propagation
time of 10.3~s, which is slightly longer than what we obtain using the
Runge-Kutta method.

\begin{figure}
\centerline{\psfig{file=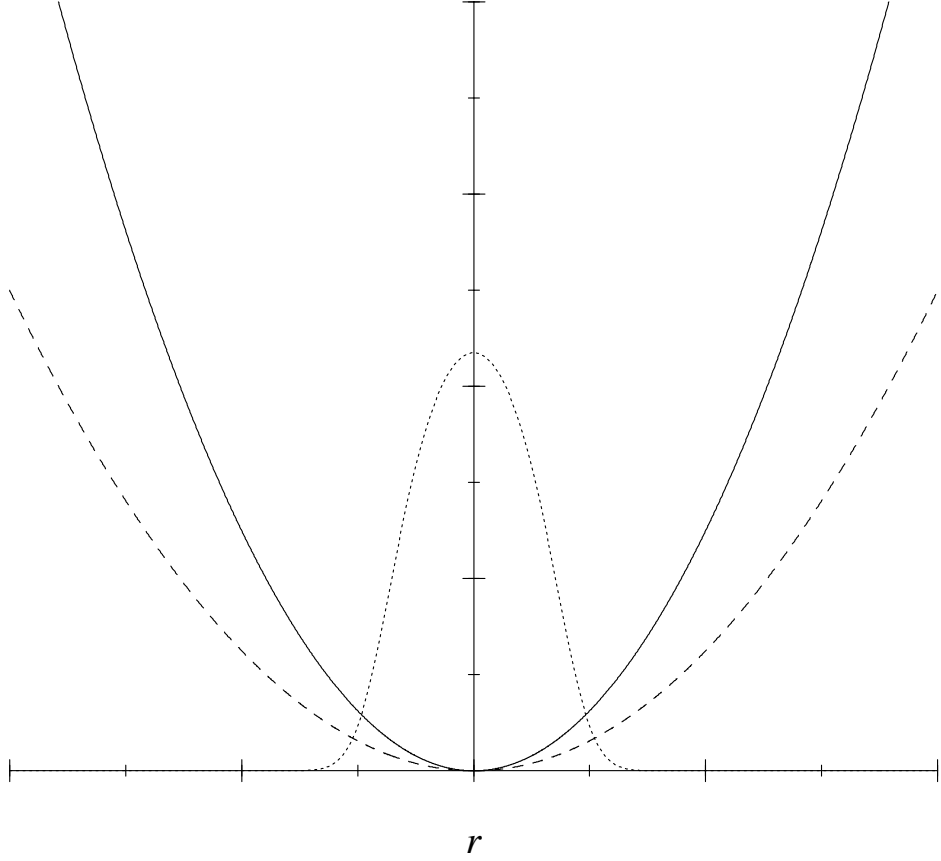,width=0.75\textwidth}}
\caption{Trapping potentials $\omega$ (solid line) and $\omega/2$
  (dashed line) used to simulate the breathing modes of a condensate.
  The wave function $\left| \psi(r) \right|^{2}$ of the stationary
  state for potential $\omega$ with $\lambda = 100$ is also given
  (dotted line).}
\label{fig:w1w2}
\end{figure}

\begin{figure}
\centerline{\psfig{file=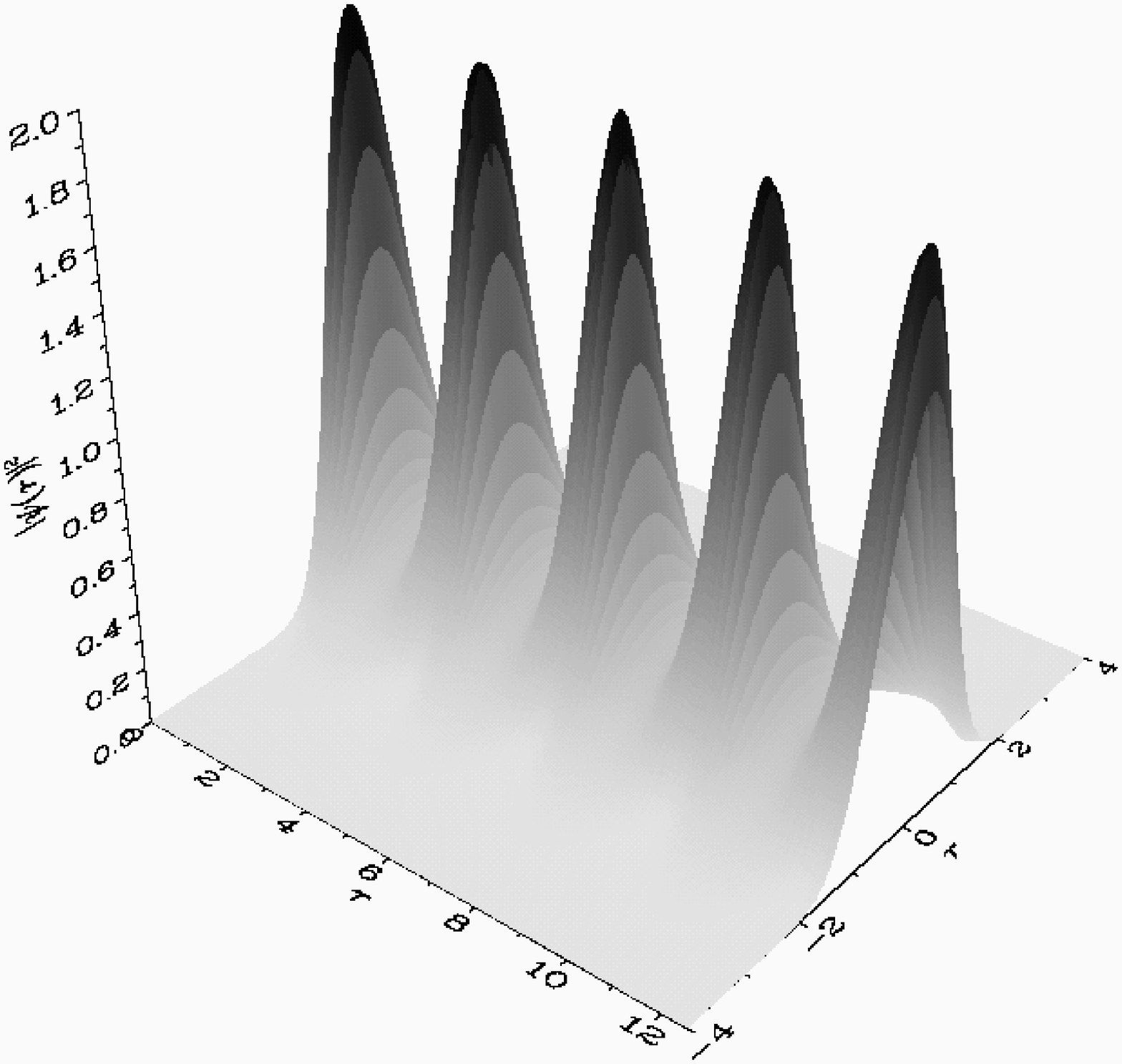,width=0.75\textwidth}}
\caption{``Breathing'' of the condensate after expansion from a trap
  of frequency $\omega$ to $\omega/2$.  The density profile $\left|
    \psi(r) \right|^{2}$ is given as a function of time $\tau$
  (scaled with respect to the final trap frequency $\omega/2$) for an initial
  $\lambda =10$ ground stationary state.}
\label{fig:breathing}
\end{figure}

We now apply our algorithm to study the dynamics of trapped
condensates.  Referring again to the spherically symmetric case, we
start with the stationary ground state for an isotropic trap frequency
$\omega$.  We then let this initial state $\psi_{0}$ evolve in a trap
of frequency $\omega/2$, as illustrated in Fig.~\ref{fig:w1w2},
corresponding to an experiment where the frequency of the potential
trapping the condensate would be instantaneously reduced by a factor
of 2.  The corresponding time-evolving wave function $\left| \psi(t,r)
\right|^{2}$ is shown in Fig.~\ref{fig:breathing}, for $\lambda=10$.
We must note that the values of $\lambda$ we give correspond to the
condensate in the initial $\omega$-frequency trap, the effective value
being used for the time evolution is thus scaled by $1/\sqrt{2}$ [see
Eq.~(\ref{eq:lambda})], while $\tau$ is rescaled with respect to the
final trap frequency $\omega/2$.  We can see the ``breathing'' of the
condensate as it expands and recontracts in the trap.

\begin{figure}
\centerline{\psfig{file=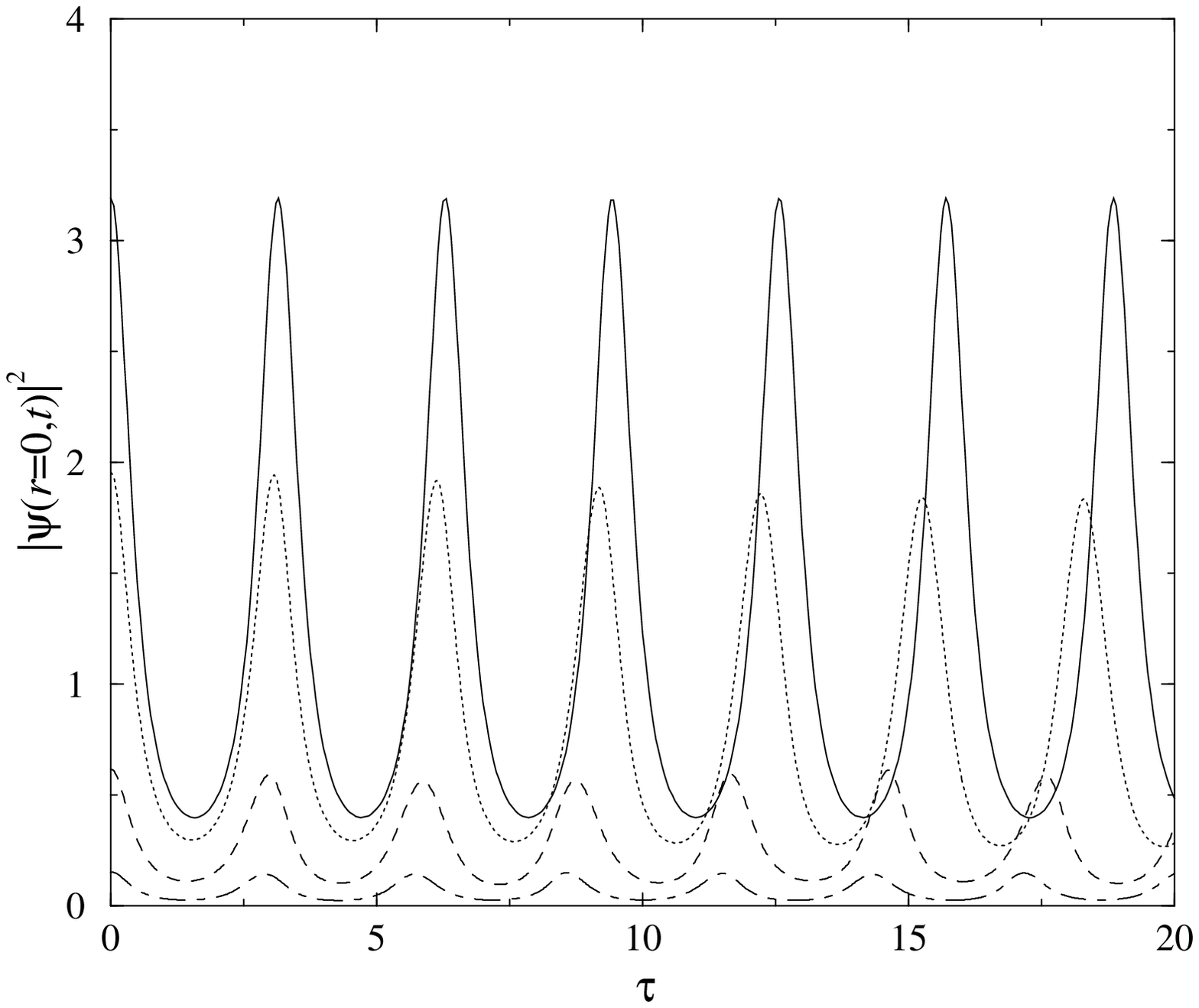,width=0.75\textwidth}}
\caption{``Breathing'' of the condensate after expansion from a trap
  of frequency $\omega$ to $\omega/2$.  The value of the wave function
  in the center of the trap, $\left| \psi(r=0) \right|^{2}$, is given
  as a function of time $\tau$ (scaled with respect to the final trap
  frequency $\omega/2$) for $\lambda$ equal to 0 (solid line), 10
  (dotted line), 100 (dashed line), and 1000 (dot-dashed line).}
\label{fig:w1w2res}
\end{figure}

It is also interesting to look at the effect of the value of the
non-linear parameter on the breathing frequency of the condensate, as
seen in Fig.~\ref{fig:w1w2res}.  
First, we note that the initial
density at the center of the trap is lower for bigger values of
$\lambda$, which is expected because of the corresponding higher
interparticle repulsion.  Starting from an unperturbed harmonic
oscillator ($\lambda=0$), for which the complete cycle time is $\tau =
4\pi$ with recurrences every $\tau = \pi$, we observe that the
oscillation frequency of the condensate in the trap increases with a
greater value of $\lambda$.

\begin{figure}
\centerline{\psfig{file=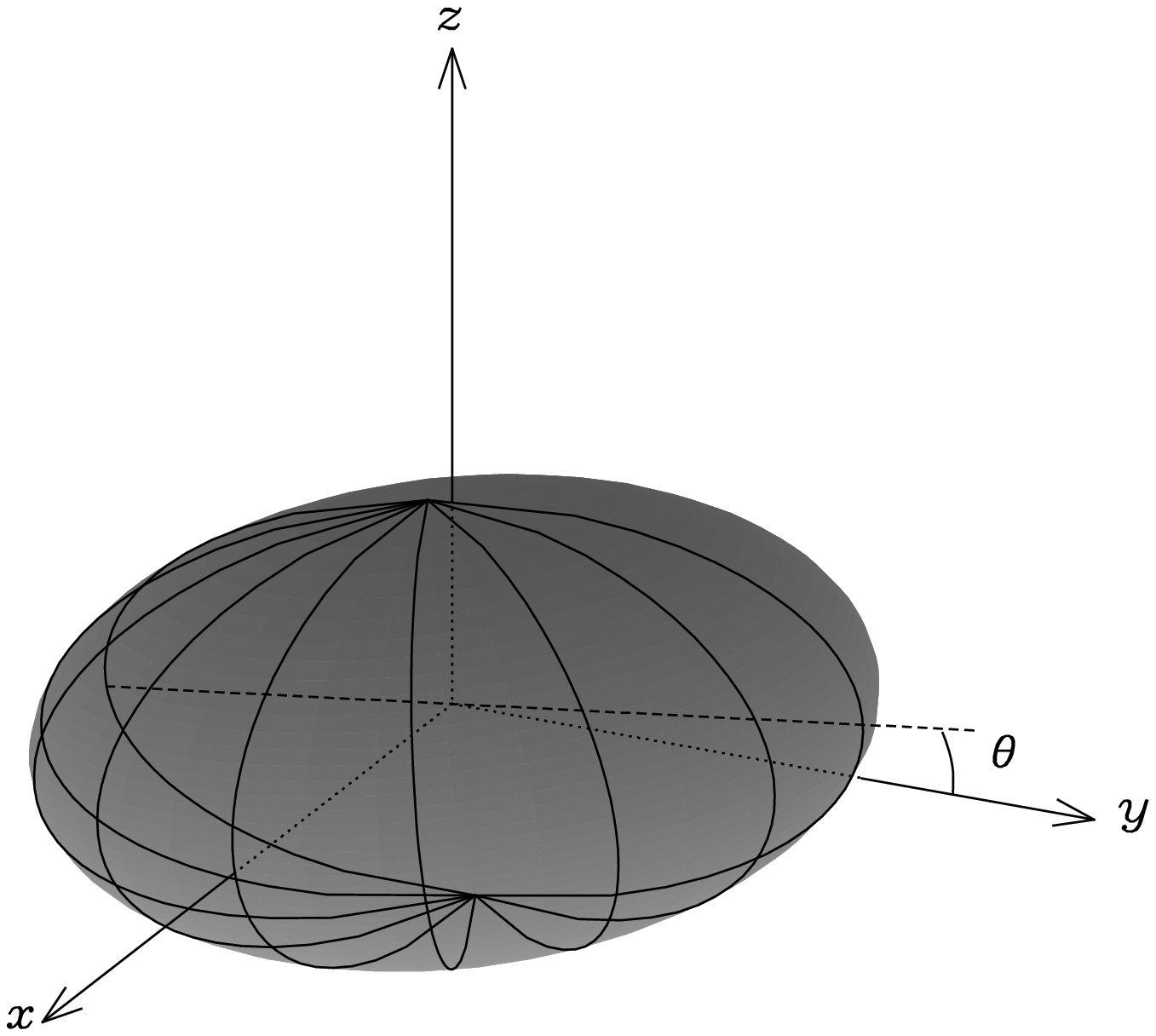,width=0.75\textwidth}}
\caption{Representation of the study a condensate's scissors mode.
  The condensate is initially tilted with respect to the trap's $y$
  and $z$ axes by an angle $\theta$.}
\label{fig:scissors}
\end{figure}

For the 3D case, we will study the scissors
mode~\cite{bec:guery-odelin99,bec:marago00} of a trapped condensate.
We consider a pancake-shaped condensate, formed in an anisotropic trap
with $\omega_{x} = \omega_{y} \ll \omega_{z}$, see
Fig.~\ref{fig:scissors}.  
The $y$ and $z$ axes of the trap are
instantaneously rotated, at $t = 0$, by angle $\theta$ around the $x$
axis.  The condensate then starts to oscillate in the trap, leading to
the so-called scissors mode.

\begin{figure}
\centerline{\psfig{file=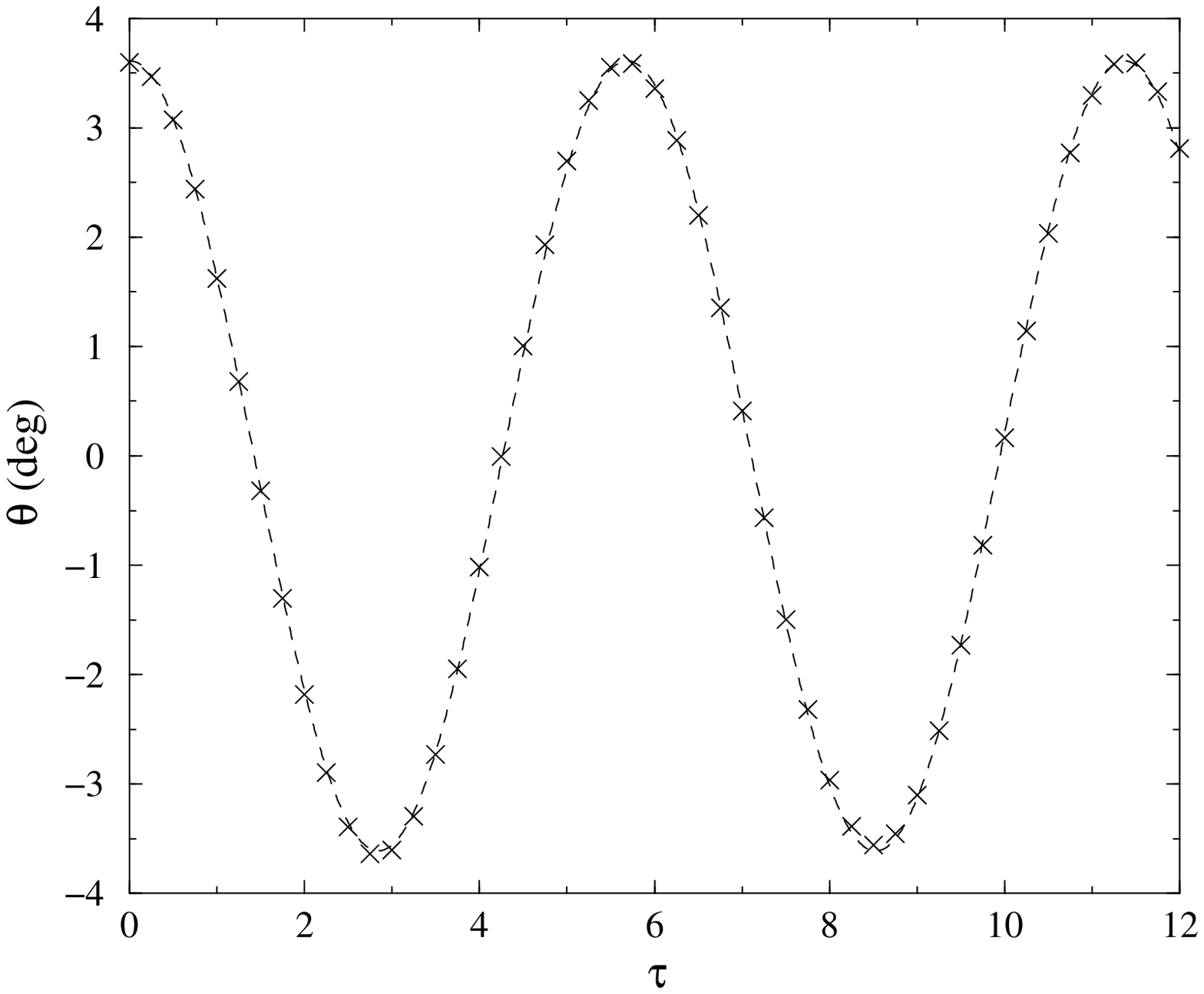,width=0.75\textwidth}}
\caption{Time evolution of the angle $\theta$ for the scissors mode.
  The crosses correspond to the angle resulting from the
  time-dependent calculation, along with the corresponding fit $\theta
  = 3.6 \cos \left(1.105 \tau\right)$ (dashed line).}
\label{fig:sc_td}
\end{figure}

Using the parameters of the experiment of Marag\`{o} \emph{et
  al.}~\cite{bec:marago00}, we first determine the stationary state
for a condensate of $N = 10^{4}$ $^{87}$Rb atoms in a trap with
$\omega_{z} = 255\ \mbox{Hz}$, $\omega_{x}/\omega_{z} =
\omega_{y}/\omega_{z} = 1/\sqrt{8}$, resulting in a value $\lambda =
147.1$.  The condensate is then tilted by an angle of $\theta =
3.6^{\circ}$, with the trapping frequency $\omega_{z}$ reduced by 2\%,
resulting in a new value of $\lambda = 148.6$.  We then calculate the
free evolution of this tilted condensate.  We report, in
Fig.~\ref{fig:sc_td}, the angle between the condensate (as determined
by the main inertia axis) and the $y$ axis, as a function of time for
the free evolution of the condensate in a trap.  The oscillation
frequency, in these conditions, is found to be $1.105$ (in rescaled
units), corresponding to 276~Hz.

This simulation was done using a basis set of $N = 29$ functions in
each dimension, using a time step $\Delta\tau = 0.005$.  The
calculation time for a propagation of duration $\tau$ is then $\approx
1735\ \mathrm{s}$.  The main advantage of using a spectral Galerkin
method, as noted in Sec.~\ref{sec:Galerkin3D}, is that in this case we
can restrict the basis set in the $x$ dimension by using only even
harmonic oscillator functions, since the reflection symmetry with
respect to the $yOz$ plane is conserved.  The number of functions is
thus reduced to $N_{x} = 15$, resulting in a decrease of CPU time to
$\approx 1030\ \mathrm{s}$ for $1\ \tau$.  This compares favorably
with the grid method, for which an equivalent calculation with $64
\times 64 \times 64$ grid points takes $\approx 1700\ \mathrm{s}$
(using the same time step and grid spacing as for the 1D grid method).

\section{Conclusion}

We have presented the application of a spectral Galerkin method to the
numerical solution of the Gross-Pitaevskii equation, describing a
Bose-Einstein condensate trapped in a harmonic potential well.  This
method is based on the decomposition of the condensate wave function
on the a basis set of eigenmodes of the harmonic oscillator, while the
nonlinear term in the GPE is calculated using the Gauss-Hermite
quadrature.  The resulting algorithm scales in $O(N^{4})$ for a full
3D problem (where $N$ is the number of basis functions used per
direction), which is slightly worse than the $O(N_p^3 \log N_p)$
scaling obtained for grid-based Fourier methods.  Nevertheless, the
required number of basis functions needed for a given problem can be
much smaller than the number of grid points $N_p$, allowing for fast
and efficient calculations using the spectral method.  We have shown
how the propagation in time can be carried out using a Runge-Kutta
method on a set of coupled ordinary differential equations.  

This method is akin to the DVR
approach~\cite{bec:schneider99a,bec:feder00}, which is relies on the
fact that matrix elements of the nonlinear term can be
\emph{approximately} evaluated to sufficiently high accuracy using an
$N$-point rule based Gauss quadratures. Our approach has the advantage
that, for the basis chosen, there are no approximations in the
computation of these integrals.  Let us however remark that the same
property can hold within the DVR method by a suitable choice of
weighted polynomials.  The main distinction between the usual DVR
approach and our method is that we treat the kinetic and potential
terms of the Hamiltonian conjointly, as detailed in
Sec.~\ref{sec:spectr}.

We have successfully applied our algorithm to simulate two different
dynamical aspects of trapped BECs.  Making use of the spherical
symmetry of an isotropic trapping potential, we used an effective 1D
equation to study the breathing of a condensate that is allowed to
expand from more confining trap to a looser one.  In the 3D case, we
have looked at the scissors modes of a pancake-shaped condensate, for
which the trapping potential is suddenly rotated along one axis.

Future work will focus on the implementation of better time-evolution
algorithms on our spectral method and on its possible parallelization.
Extensions will also be made to consider other terms in the
Gross-Pitaevskii Hamiltonian, such as the potential created by the
interaction with a laser field, or coupled Gross-Pitaevskii equations
used in the simulation of two-species
condensates~\cite{bec:trippenbach00} or of the formation of molecules
in atomic condensates~\cite{bec:timmermans99,bec:heinzen00}.

\begin{acknowledgments}
We thank L. DiMenza, O. Dulieu, J. Mary, F. Masnou-Seeuws, and P. Pellegrini
for stimulating discussions.  Part of this work was carried out using
the computer resources of IDRIS (Orsay, France).
\end{acknowledgments}

\end{document}